\documentclass[prb,superscriptaddress,amsmath,amssymb,floatfix,margin=1in,super,twocolumn,longbibliography]{revtex4-1} 

\usepackage[dvips]{graphicx}
\usepackage{xspace}
\usepackage[usenames,dvipsnames]{xcolor}
\usepackage{array}
\usepackage{ulem}
\usepackage{hyperref}
\usepackage[english]{babel}
\usepackage{csquotes}
\usepackage{multirow}
\usepackage{dblfloatfix}

\setcitestyle{super}

\newcommand{\COSO}{Cu$_{2}$OSeO$_{3}$\@\xspace}

\newcommand{\phii}{Institute of Physics 2, Faculty of
Mathematics and Natural Sciences, University of Cologne, Z\"{u}lpicher Stra{\ss}e 77, D-50937 Cologne, Germany}
\newcommand{\mineralogy}{Institute of Geology and Mineralogy, Faculty of
Mathematics and Natural Sciences, University of Cologne, Z\"{u}lpicher Stra{\ss}e 49b, D-50674 Cologne, Germany}

\begin{document}

\title{Inelastic light scattering from optical magnons and localized spin excitations in the cluster magnet Cu$_{2}$OSeO$_{3}$}
\title{Magnetic and phononic inelastic light scattering in the cluster magnet Cu$_{2}$OSeO$_{3}$}
\title{Inelastic light scattering in the spin cluster Mott insulator Cu$_{2}$OSeO$_{3}$}
\author{R.~B.~Versteeg}
 \email[Corresponding author:]{versteeg@ph2.uni-koeln.de}
\author{C.~Boguschewski}
  \affiliation{\phii} 
\author{P.~Becker}  
\affiliation{\mineralogy}  
\author{P.~H.~M.~van Loosdrecht} 
 \email[Corresponding author:]{pvl@ph2.uni-koeln.de}
  \affiliation{\phii}

\date{\today}

\begin{abstract}

Clusters of single spins form the relevant spin entities in the formation of long-range magnetic order in spin cluster Mott insulators. Such type of spin order bears resemblance to molecular crystals, and we therefore may expect a prototypical spin wave spectrum which can be divided into low-energy external and high-energy internal cluster spin wave modes. Here, we study high-energy spin cluster excitations in the spin cluster Mott insulator Cu$_{2}$OSeO$_{3}$ by means of spontaneous Raman scattering. Multiple high-energy optical magnon modes are observed, of which the Raman-activity is shown to originate in the Elliot-Loudon scattering mechanism. Upon crossing the long-range ordering transition temperature the magnetic modes significantly broaden, corresponding to scattering from localized spin excitations within the spin clusters. Different optical phonon modes show a strong temperature dependence, evidencing a strong magnetoelectric coupling between optical phonons and the high-energy spin cluster excitations. Our results support the picture that \COSO can be regarded as a solid-state molecular crystal of spin nature. 

\end{abstract}

\maketitle

\section{Introduction}

Strong electron-electron correlations lie at the origin of the formation of a vast range of exotic charge, orbital, and spin states in solids.\cite{patrik1999lecture,khomskii2014transition} For the majority of quantum materials, the single ionic site spin, orbitals, and charge form the entities for the description of the ordering and resulting collective excitations. However, in a peculiar subclass of quantum materials, known as cluster Mott-insulators, \cite{attfield2015orbital,streltsov2017orbital} the single site description does not hold, and instead \enquote{molecules} (or clusters) of spin, orbital, and charge degree of freedom form the relevant entities to describe the emerging physical phenomena. This solid-state molecule formation has, for instance, been found to underlie the nature and speed limit of the Verwey-transition in magnetite, \cite{senn2012charge,dejong2013speed} and allowed to demonstrate Young-interference in the resonant inelastic x-ray scattering process from iridate dimer molecules.\cite{revelli2019resonant} 
    
The solid-state molecule formation may be fluctuating, as in the case of dimer formation in the resonating valence bond condensated state,\cite{kimber2014,anderson1987resonating} but we can also identify \enquote{rigid} molecular crystals of charge, orbital and spin degree of freedom. The latter situation occurs in materials with a disproportionation in structural bond lengths,\cite{kimber2014} with a resulting subdivision of strong and weak electronic interactions. This situation leads to a collective excitation spectrum which can be subdivided into low-energy \textit{external} and high-energy \textit{internal} modes, in close resemblance to the vibrational spectrum of the true molecular crystal.\cite{venkataraman1970,natkaniec1980phonon} An understanding of the collective excitation spectrum below and above the \enquote{crystallization} temperature is important as it provides the dynamic fingerprint of the solid-state molecular crystal nature of cluster Mott insulators. 
    
In this context, we investigate the temperature dependent inelastic light scattering response of the spin cluster Mott insulator \COSO. While this material was initially of high interest because of the skyrmion metamagnetism,\cite{seki2012,longwavelength,seki2012formation} a second surprise came with the insight that effective $S$\,=\,$1$ Cu$_4$ spin clusters form the relevant spin entities for the formation of long-range order, instead of the single site Cu$^{2+}$ $S$\,=\,$\tfrac{1}{2}$ spins.\cite{clustercoso,romhanyi2014} This picture was consecutively firmly established with electron spin resonance and inelastic neutron scattering studies deep inside the ordered phase, which showed that \COSO has a characteristic spin wave spectrum comprising of low-energy cluster-external and high-energy cluster-internal modes.\cite{ozerov2014,portnichenko2016,tucker2016} The spin cluster excitation nature in the paramagnetic phase has only been minimally discussed.\cite{laurita2017,versteeg2019} A multitude of the high-energy spin cluster excitations have been observed in different Raman studies, \cite{gnezdilov2010,kurnosov2012,versteeg2019} and used in a time-resolved Raman study to track photoinduced spin cluster disordering,\cite{versteeg2019} but insight in the underlying magnetic light scattering mechanism is still incomplete. We address these issues in greater detail in this article. 
    
With spontaneous Raman-spectroscopy we observed multiple high-energy spin excitations in \COSO, which can be assigned to different spin cluster transitions. Their Raman-activity can be traced back to the Elliot-Loudon scattering mechanism. While the spin cluster excitations correspond to well-defined optical magnons in the long-range ordered phase, they cross over into localized cluster-internal spin excitations above $T_{\rm C}$, resulting in a broad magnetic scattering continuum. Different optical phonon modes show a strong temperature dependence, evidencing a strong magnetoelectric coupling between optical phonons and the high-energy spin cluster excitations. Our results support the picture that \COSO can be regarded as a solid-state molecular crystal of spin nature. 

\vspace*{\fill}
    
\section{Spin cluster formation}
\begin{figure}[h!]
 \center
\includegraphics[width=3.375in]{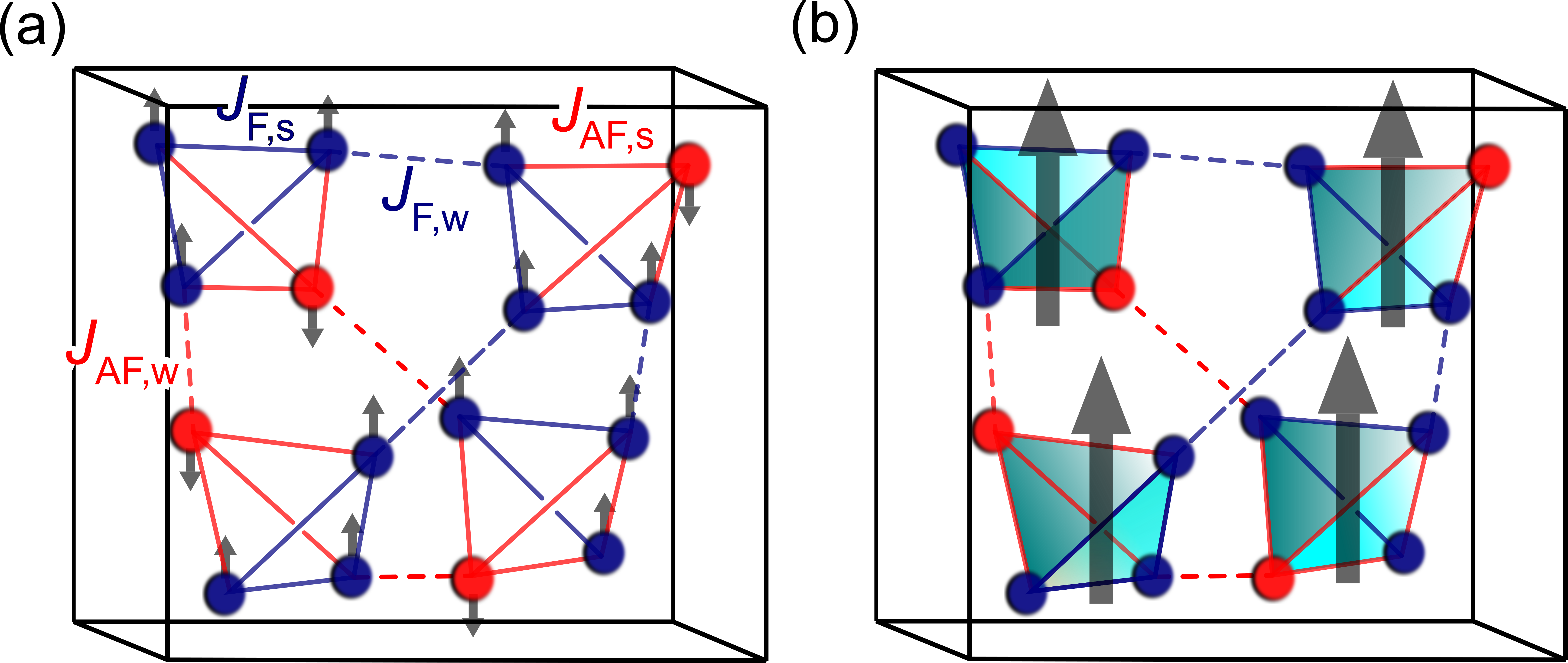}    
\caption{(a) The magnetic unit cell of \COSO. A few exchange interactions are indicated as $J_{\rm F,s}$, $J_{\rm AF,s}$, $J_{\rm F,w}$, and $J_{\rm AF,w}$ where the subscripts refer to (anti)ferromagnetic (AF/F) and strong or weak exchange (s/w). The blue lines are predominantly ferromagnetic exchange couplings and the red lines predominantly antiferromagnetic exchange couplings. The full lines are strong exchange couplings and the dashed lines weak exchange couplings. Not shown is the antiferromagnetic exchange path $J_{\rm AF,OO}$ which couples spins across a hexagon of alternating Cu-I and Cu-II sites. (b) The strong exchange interaction leads to effective $S$\,=\,$1$ spin clusters.}
\label{fig:clustercoso}
\end{figure}

Before turning to the Raman spectroscopy results, we summarize the most important conclusions from the works Refs.\,\citenum{clustercoso,romhanyi2014,ozerov2014,tucker2016,portnichenko2016}, which discuss the spin cluster order and excitations in \COSO in considerable detail. This summary will be beneficial in order to assign the Raman-active cluster modes and deduce the inelastic light scattering mechanism. Figure \ref{fig:clustercoso}a shows the magnetic unit cell of \COSO. The localized Cu$^{2+}$ $S$\,=\,$1/2$ spins reside on the vertices of corned-shared tetrahedra in a distorted pyrochlore lattice. \cite{clustercoso} DFT+$U$ calculations reveal that these tetrahedra can be separated into tetrahedra of \enquote{stronger} and \enquote{weaker} exchange energy scales.\cite{clustercoso,ozerov2014} A few exchange couplings are indicated in Fig.\,\ref{fig:clustercoso}a as $J_{\rm F,s}$, $J_{\rm AF,s}$, $J_{\rm F,w}$, and $J_{\rm AF,w}$, where the subscripts refer to (anti)ferromagnetic (AF/F) and strong or weak exchange (s/w). 
An additional antiferromagnetic exchange $J_{\rm AF,OO}$ couples spins across a hexagon of alternating Cu-I and Cu-II sites (not shown).\cite{footnote1} The inversion symmetry between spin is absent both inside the cluster and in between clusters.\cite{jwbos2008} This results in a nonzero Dzyaloshinskii-Moriya-interaction $D$ across all drawn paths, with a $D/J$-ratio on the order of $0.1$\,$-$\,$0.6$ for different Cu$^{2+}$-Cu$^{2+}$ bonds.\cite{clustercoso,yang2012}  

The strong Heisenberg (and Dzyaloshinskii-Moriya exchange) interactions couple four localized $S$\,=\,$1/2$ spins into a three-up-one-down $S$\,=\,$1$ entity. Here the Cu-II ions couple ferromagnetically through $J_{\rm F,s}$, while the Cu-I ion couples antiferromagnetically to the Cu-II ions through $J_{\rm AF,s}$. The Cu$_4$ cluster formation occurs far above the long-range ordering temperature of $T_{\rm C}$\,$\approx$\,$58$\,K, as evidenced from the temperature dependence of the inverse magnetic susceptibility, which in the paramagnetic phase shows gradual crossover between a $S$\,=\,$1/2$ Curie-constant at high temperatures and a $S$\,=\,$1$ Curie-constant at lower temperatures.\cite{tucker2016} The Curie-constant crossover is characteristic of materials with strong and weak exchange exchange interactions. \cite{samuelsen1973} 

For a single isolated cluster the relevant exchange paths are the strong paths $J_{\rm F,s}$ and $J_{\rm AF,s}$. The spin Hamiltonian for the isolated cluster is thus given by:

\begin{equation}
\begin{split}
\hat{\mathcal{H}}_0 & = J_{\rm AF,s}\hat{S}_1\cdot \big(\hat{S}_{2}+\hat{S}_{3}+\hat{S}_{4}\big) + \\ 
& J_{\rm FM,s}\big( \hat{S}_2\cdot\hat{S}_{3} + \hat{S}_3\cdot\hat{S}_{4} + \hat{S}_4\cdot\hat{S}_{2 } \big) 
\end{split}
\end{equation}

Here $\hat{S}_1$ is the spin angular momentum operator for the spin on the Cu-I ion, and $\hat{S}_2$, $\hat{S}_3$, and $\hat{S}_4$ for the Cu-II ions. The eigenstates are notated as $|S,S^z \rangle_{\rm R}$. Here $S$ indicates the total spin quantum number of the cluster. The secondary spin quantum number (the spin momentum projection along the z-axis) is indicated by $S^z$. The symmetry label ${\rm R}$ refers to the irreducible representations of the single cluster's $C_{3v}$ point group (this is the cluster symmetry when only the magnetic Cu$^{2+}$ ions are considered). Under the symmetry $C_{3v}$ the $2^4$\,$=$\,$16$-dimensional Hilbert space of a single tetrahedron splits into the ground state A$_1$-triplet $|1,S^z \rangle_{\rm A_1}$, and two E$_1$ and E$_2$ singlets, one A$_1$ quintet $|2,S^z \rangle_{\rm A_1}$ , and two E$_1$ and E$_2$ triplets excited states $|1,S^z \rangle_{\rm E_1/E_2}$.\cite{romhanyi2014} The corresponding (isolated) spin cluster wave functions are indicated in Table \ref{table:wavefunctions}, and will be later of use when discussing the Raman-activity of the cluster excitations. From the form of the wave functions it becomes apparent that the cluster wave functions are highly entangled. 

The interaction between clusters is to first approximation captured by the tetrahedral mean field (TMF) Hamiltonian:

\begin{equation}
\hat{\mathcal{H}}_{\rm TMF}  = \hat{\mathcal{H}}_0 +  \hat{\mathcal{H}}'\big[\langle \hat{S}_1 \rangle, \langle \hat{S}_{2,3,4} \rangle, J_{\rm FM,w}, J_{\rm AF,w}, J_{\rm AF,OO}\big]
\end{equation}

The perturbation term $\hat{\mathcal{H}}'$ depends on the mean magnetic fields exerted by the Cu-I ions and Cu-II ions, which are proportional to the magnetic moments $\langle \hat{S}_1 \rangle$, and $ \langle \hat{S}_{2,3,4} \rangle$ respectively, the weak inter-cluster exchange couplings  $J_{\rm FM,w}$ and $J_{\rm AF,w}$, and a hexagonal antiferromagnetic weak exchange path $J_{\rm AF,OO}$.\cite{footnote1} A further refinement to the model was done in the second quantization formalism. \cite{romhanyi2014} The resulting energies of the cluster states at the $\Gamma$-point with corresponding degeneracies are indicated in table 2. 

As a result of the inter-cluster interactions the excited state quintet furthermore mixes into the ground state of the isolated cluster model:

\begin{equation}
|g_{\rm TMF} \rangle = \cos \tfrac{\alpha}{2}|1,1 \rangle_{\rm A_1} + \sin \tfrac{\alpha}{2}|2,1 \rangle_{\rm A_1}
\label{eq:groundstate}
\end{equation}

Here $S^z$\,=\,$+1$ has been chosen as the ground state. Note that in the interacting model the total spin quantum number $S$ isn't a good quantum number anymore. The factor $\tfrac{\alpha}{2}$ gives the amount of quintet mixing, for which $\alpha$\,$\approx$\,$0.58$ was found. \cite{romhanyi2014} A possible perturbation of the excited state wave functions by the inter-tetrahedral interaction is not discussed in the aforementioned papers. In the later discussion of the Raman spectra we therefore use the perturbed ground state $|g_{\rm TMF}\rangle$ instead of $\vert 1\rangle$, but use the unperturbed single cluster wave functions $\vert 2\rangle$ to $\vert 16\rangle$ for the excited states. These prove to be sufficient to explain all observed Raman modes of magnetic origin.

\begin{table*}
\begin{center}
\caption{The 16 isolated single cluster wave functions. The state notation is indicated by $|S,S^z \rangle_{\rm R}$. The superposition for the different cluster wave functions are fully written out. $\vert 1,1\rangle_{\rm A_1}$ gives the single cluster ground state. The ground state $|g_{\rm TMF}\rangle$ in the interacting cluster model is a superposition of the states $\vert 1,1\rangle_{\rm A_1}$ and \textbf{$\vert 2,1\rangle_{\rm A_1}$}. }
\def\arraystretch{1.25}
\small
\begin{tabular}{|c|c|c|c|}
\hline 
N & $\vert n\rangle$ & $\vert S,S^z\rangle$ & full wavefunction\tabularnewline
\hline 
\hline 
\multirow{3}{*}{0} & $\vert 1\rangle$ & $\vert 1,1\rangle_{\rm A_1}$ & $\frac{1}{2\sqrt{3}}\big(3\vert\downarrow\uparrow\uparrow\uparrow\rangle-\vert\uparrow\downarrow\uparrow\uparrow\rangle-\vert\uparrow\uparrow\downarrow\uparrow\rangle-\vert\uparrow\uparrow\uparrow\downarrow\rangle\big)$\tabularnewline
\cline{2-4} 
 & $\vert 2\rangle$ & $\vert 1,0\rangle_{\rm A_1}$ & $\frac{1}{\sqrt{6}}\big(\vert\downarrow\downarrow\uparrow\uparrow\rangle +\vert\downarrow\uparrow\downarrow\uparrow\rangle+\vert\downarrow\uparrow\uparrow\downarrow\rangle-\vert\uparrow\downarrow\downarrow\uparrow\rangle-\vert\uparrow\downarrow\uparrow\downarrow\rangle-\vert\uparrow\uparrow\downarrow\downarrow\rangle\big)$\tabularnewline
\cline{2-4} 
 & $\vert 3\rangle$ & $\vert 1,\bar{1}\rangle_{\rm A_1}$ & $\frac{1}{2\sqrt{3}}\big(\vert\downarrow\downarrow\downarrow\uparrow\rangle +\vert\downarrow\downarrow\uparrow\downarrow\rangle +\vert\downarrow\uparrow\downarrow\downarrow\rangle -3\vert\uparrow\downarrow\downarrow\downarrow\rangle\big)$\tabularnewline
\hline 
\multirow{2}{*}{1} & $\vert 4\rangle$ & $\vert 0,0\rangle_{\rm E_1}$ & $\frac{1}{2\sqrt{3}}\big(2\vert\downarrow\downarrow\uparrow\uparrow\rangle-\vert\downarrow\uparrow\downarrow\uparrow\rangle-\vert\downarrow\uparrow\uparrow\downarrow\rangle-\vert\uparrow\downarrow\downarrow\uparrow\rangle-\vert\uparrow\downarrow\uparrow\downarrow\rangle+2\vert\uparrow\uparrow\downarrow\downarrow\rangle\big)$\tabularnewline
\cline{2-4} 
 & $\vert 5\rangle$ & $\vert 0,0\rangle_{\rm E_2}$ & $\frac{1}{2}\big(\vert\downarrow\uparrow\downarrow\uparrow\rangle-\vert\downarrow\uparrow\uparrow\downarrow\rangle-\vert\uparrow\downarrow\downarrow\uparrow\rangle+\vert\uparrow\downarrow\uparrow\downarrow\rangle\big)$\tabularnewline
\hline 
\multirow{5}{*}{2} & $\vert 6\rangle$ & $\vert 2,\bar{2}\rangle_{\rm A_1}$ & $\vert\downarrow\downarrow\downarrow\downarrow\rangle$\tabularnewline
\cline{2-4} 
 & $\vert 7\rangle$ & $\vert 2,\bar{1}\rangle_{\rm A_1}$ & $\frac{1}{2}\big(\vert\downarrow\downarrow\downarrow\uparrow\rangle+\vert\downarrow\downarrow\uparrow\downarrow\rangle+\vert\downarrow\uparrow\downarrow\downarrow\rangle+\vert\uparrow\downarrow\downarrow\downarrow\rangle\big)$\tabularnewline
\cline{2-4} 
 & $\vert 8\rangle$ & $\vert 2,0\rangle_{\rm A_1}$ & $\frac{1}{\sqrt{6}}\big(\vert\downarrow\downarrow\uparrow\uparrow\rangle+\vert\downarrow\uparrow\downarrow\uparrow\rangle+\vert\downarrow\uparrow\uparrow\downarrow\rangle+\vert\uparrow\downarrow\downarrow\uparrow\rangle+\vert\uparrow\downarrow\uparrow\downarrow\rangle+\vert\uparrow\uparrow\downarrow\downarrow\rangle\big)$\tabularnewline
\cline{2-4} 
 & $\vert 9\rangle$ & $\vert 2,1\rangle_{\rm A_1}$ & $\frac{1}{2}\big(\vert\downarrow\uparrow\uparrow\uparrow\rangle+\vert\uparrow\downarrow\uparrow\uparrow\rangle+\vert\uparrow\uparrow\downarrow\uparrow\rangle+\vert\uparrow\uparrow\uparrow\downarrow\rangle\big)$\tabularnewline
\cline{2-4} 
 & $\vert 10\rangle$ & $\vert 2,2\rangle_{\rm A_1}$ & $\vert\uparrow\uparrow\uparrow\uparrow\rangle$\tabularnewline
\hline 
\multirow{6}{*}{3} & $\vert 11\rangle$ & $\vert 1,\bar{1}\rangle_{\rm E_1}$ & $\frac{1}{\sqrt{6}}\big(\vert\downarrow\downarrow\downarrow\uparrow\rangle+\vert\downarrow\downarrow\uparrow\downarrow\rangle-2\vert\downarrow\uparrow\downarrow\downarrow\rangle\big)$\tabularnewline
\cline{2-4} 
 & $\vert 12\rangle$ & $\vert 1,\bar{1}\rangle_{\rm E_2}$ & $\frac{1}{2}\big(\vert\downarrow\downarrow\downarrow\uparrow\rangle-\vert\downarrow\downarrow\uparrow\downarrow\rangle\big)$\tabularnewline
\cline{2-4} 
 & $\vert 13\rangle$ & $\vert 1,0\rangle_{\rm E_1}$ & $\frac{1}{2\sqrt{3}}\big(-2\vert\downarrow\downarrow\uparrow\uparrow\rangle+\vert\downarrow\uparrow\downarrow\uparrow\rangle+\vert\downarrow\uparrow\uparrow\downarrow\rangle-\vert\uparrow\downarrow\downarrow\uparrow\rangle-\vert\uparrow\downarrow\uparrow\downarrow\rangle+2\vert\uparrow\uparrow\downarrow\downarrow\rangle\big)$\tabularnewline
\cline{2-4} 
 & $\vert 14\rangle$ & $\vert 1,0\rangle_{\rm E_2}$ & $\frac{1}{2}\big(\vert\downarrow\uparrow\downarrow\uparrow\rangle-\vert\downarrow\uparrow\uparrow\downarrow\rangle+\vert\uparrow\downarrow\downarrow\uparrow\rangle-\vert\uparrow\downarrow\uparrow\downarrow\rangle\big)$\tabularnewline
\cline{2-4} 
 & $\vert 15\rangle$ & $\vert 1,1\rangle_{\rm E_1}$ & $\frac{1}{\sqrt{6}}\big(-2\vert\uparrow\downarrow\uparrow\uparrow\rangle+\vert\uparrow\uparrow\uparrow\downarrow\rangle+\vert\uparrow\uparrow\uparrow\downarrow\rangle\big)$\tabularnewline
\cline{2-4} 
 & $\vert 16\rangle$ & $\vert 1,1\rangle_{\rm E_2}$ & $\frac{1}{\sqrt{2}}\big(\vert\uparrow\uparrow\downarrow\uparrow\rangle-\vert\uparrow\uparrow\uparrow\downarrow\rangle\big)$\tabularnewline
\hline 
\end{tabular}
\label{table:wavefunctions}
\end{center}
\end{table*}

\section{Experimental details}
\subsection{Sample preparation}
Single crystals of \COSO were synthesized by chemical transport reaction growth. Stoichiometric amounts of CuO and SeO$_2$ powders (both ChemPur, 99.999\%), with an addition of TeCl$_4$ (Sigma Aldrich, 99.999\%) as transporting agent, were sealed in evacuated SiO$_2$-glass ampoules. The ampoules were placed in horizontal two-zone tube furnaces, and heated to $893$\,K at the source side, and $773$\,K at the sink side of the ampoules. After a growth period of circa $40$ days dark green crystals of several $\sim$\,mm$^3$ size, with well-developed \{100\}, \{110\} and \{111\} morphological faces resulted. For the Raman study a (111) oriented plate-shaped sample was prepared with a flat as-grown (111) face, and a lapped parallel opposite surface, polished with $1$\,$\mu$m  grit size diamond paste. 

\subsection{Spontaneous Raman spectroscopy}
The Raman scattering experiments were performed at low temperatures ranging from $7.5$\,K to $150$\,K. The sample is placed in a \textsc{Oxford Microstat} with a temperature stability of $0.1$\,K. The used excitation light is provided by a frequency doubled Nd:YAG  (central wavelength $\lambda_C$=$532$\,nm) laser. The polarization for the excitation light is cleaned with a Glan Taylor polarizer. The scattered light polarization is analyzed with a sheet polarizer. We used a confocal backscattering geometry, with a NA\,=\,$0.5$ microscope objective which illuminates the sample and collects the scattered light. The excitation density was kept below $500$\,W/cm$^{2}$. Laser heating effects are minimal since the $532$\,nm excitation falls within the transmission window of \COSO (Ref.\,\citenum{versteeg2016}) A \textsc{Jobin Yvon T64000} triple subtractive spectrometer, equipped with a \textsc{Symphony} $1024$\,$\times$\,$256$ charge-coupled device, was used to detect the scattered light. The resolution in the studied energy interval lies below $2$\,cm$^{-1}$. Porto notation z(x,y)$\bar{z}$ is used to indicate the polarization of the incoming (x) and scattered (y) light, with the light wave vector parallel to z. The x-polarization lies along the crystallographic [1$\bar{1}$0] axis, and y-polarization along [11$\bar{2}$].

\section{Phonon scattering}
The noncentrosymmetric cubic lattice of \COSO is described by space group $P2_13$. There are $Z$\,$=$\,$8$ chemical formula units in the structural unit cell (16 Cu$^{2+}$-atoms in total). This gives a total of $7$\,$\times$\,$8$\,$\times$\,$3$\,$=$\,$168$ phonons. There are 5 atoms on a $4a$ Wyckoff-position, and 3 atoms on a $12b$ Wyckoff-position. \cite{jwbos2008,gnezdilov2010}

The total $\Gamma$-point phonon spectrum is decomposed in the following irreducible representations:

\begin{equation}
\Gamma = 14{\rm A} + 14{\rm E}_1 +14{\rm E}_2 + 42{\rm T}    
\end{equation}

\noindent where the acoustic phonons contribute 1T. \cite{POINT} 

The $\Gamma$-point optical phonon spectrum is decomposed in:

\begin{equation}
\Gamma^{\rm optical} = 14{\rm A}^{\rm (R)} + 14{\rm E}_1^{\rm (R)} +14{\rm E}_2^{\rm (R)} + 41{\rm T}^{\rm (R,IR)}    
\end{equation}
The threefold degenerate T-irrep phonons are Raman (R) and infrared active (IR), whereas the one-fold degenerate A, E$_1$, and E$_2$ are only Raman-active (R).
For the (111) oriented sample the A Raman modes will only show up in parallel polarization geometry, while the E$_1$, E$_2$, and T modes are observable in both parallel and crossed polarization geometry. \cite{gnezdilov2010}

\begin{figure*}
 \center
\includegraphics[scale=1]{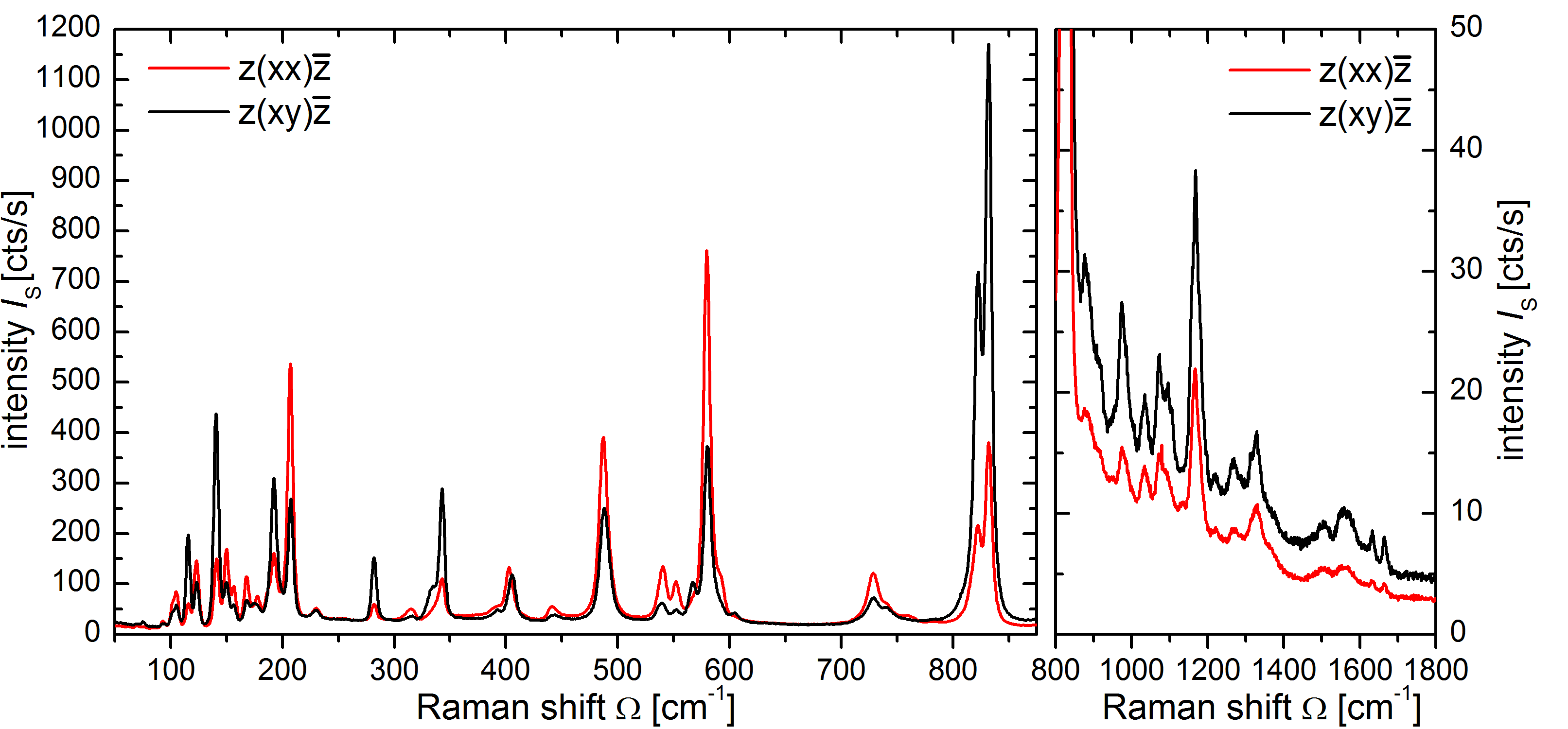} 
\caption{Phonon spectrum obtained at $T$\,=$80$\,K for parallel z(x,x)$\bar{z}$ and perpendicular z(x,y)$\bar{z}$ polarization geometries. Strong phonon modes are observed in the frequency range up to $850$cm$^{-1}$. Between $850$cm$^{-1}$ and $1800$cm$^{-1}$ weaker phonon modes are observed.}
\label{fig:phonons}
\end{figure*}

\twocolumngrid

Figure \ref{fig:phonons} shows the phonon spectrum in z(x,x)$\bar{z}$ and z(x,y)$\bar{z}$ polarization configuration at $T$\,=\,$80$\,K over the range $50$-$1800$cm$^{-1}$. The spectra agree with the observations of Gnezdilov \textit{et al.} (Ref.\,\citenum{gnezdilov2010}), which reports the observation of $53$ strong optical phonons in the frequency range up to $850$\,cm$^{-1}$ and $21$ weak optical phonons in the frequency range between $850-2000$\,cm$^{-1}$. For completeness, $26$ $T$-phonons were detected in the infrared absorption spectrum by Miller \textit{et al.} (Ref.\,\citenum{miller2010}), where also the nature of the phonons is thoroughly discussed.

\clearpage

\section{Spin cluster excitation scattering}
\subsection{Mode assignment and identification of scattering mechanism}
\begin{figure}[h!]
 \center
\includegraphics[scale=1]{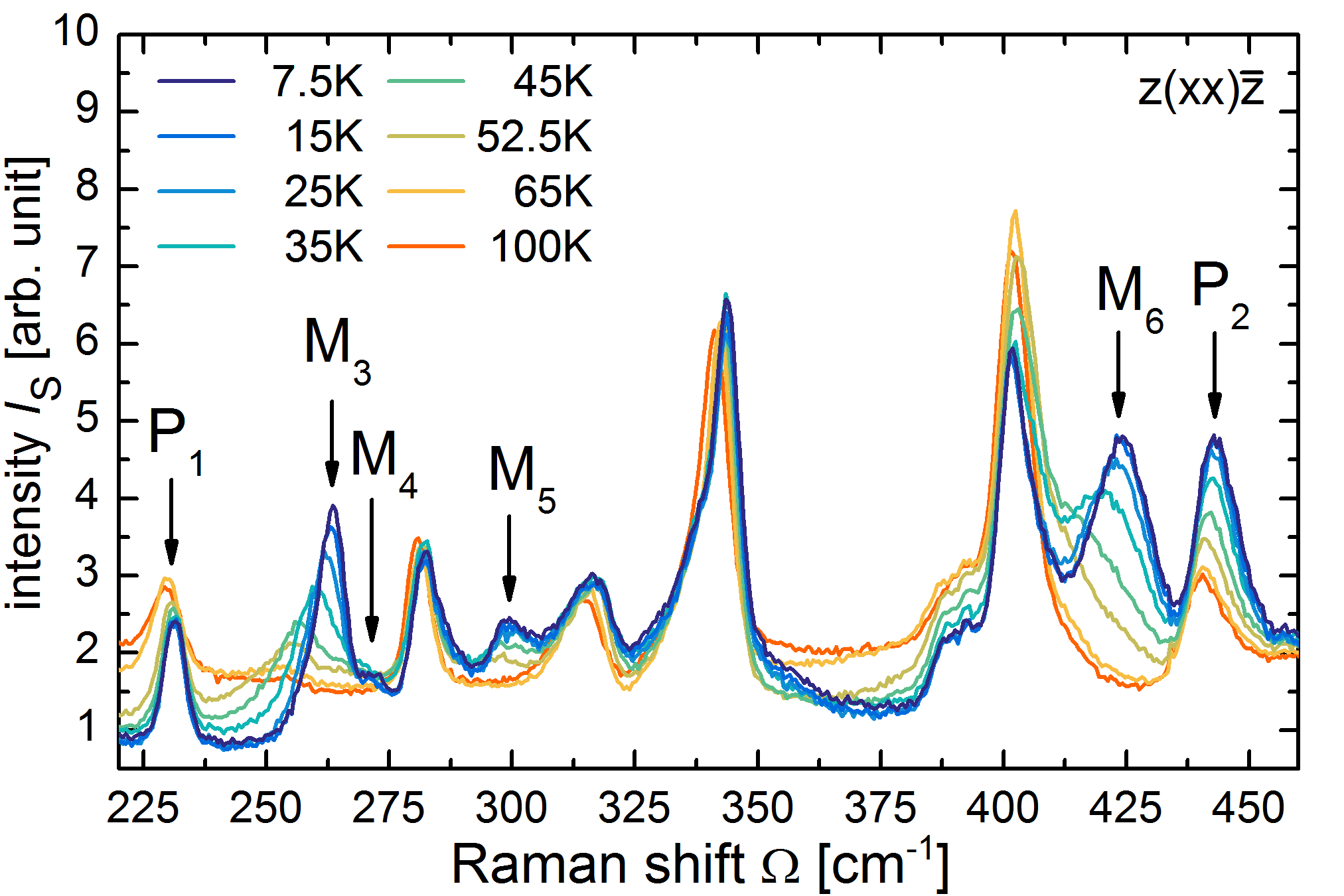} 
\caption{Normalized temperature dependent Raman spectra for \COSO in z(x,x)$\bar{z}$ polarization configuration. Four high-energy spin cluster excitations are identified within this energy range: $263$\,cm$^{-1}$ (M$_3$), $273$\,cm$^{-1}$ (M$_4$), $300$\,cm$^{-1}$ (M$_5$), and $425$\,cm$^{-1}$ (M$_6$). Two phonon modes of interest are indicated with P$_1$ ($231$\,cm$^{-1}$) and P$_2$ ($444$\,cm$^{-1}$).}
\label{fig:magnonraman}
\end{figure}

Figure \ref{fig:magnonraman} shows temperature dependent Raman spectra in the range $220-460$\,cm$^{-1}$ for the z(x,x)$\bar{z}$ polarization configuration. The spectra have been normalized to the phononic scattering intensity in the region $520-610$\,cm$^{-1}$. In this energy range multiple strongly scattering modes of magnetic origin are identified: $263$\,cm$^{-1}$ (M$_{3}$), $273$\,cm$^{-1}$ (M$_4$), $300$\,cm$^{-1}$ (M$_5$), and $425$\,cm$^{-1}$ (M$_6$). The two weak magnetic modes at $86$\,cm$^{-1}$ (M$_1$) and $204$\,cm$^{-1}$ (M$_2$) are not shown.\cite{gnezdilov2010,kurnosov2012} Two phonon modes of interest are indicated with P$_1$ ($231$\,cm$^{-1}$) and P$_2$ ($444$\,cm$^{-1}$).

We wish to discuss the possible cluster transitions from the ground state $|g_{\rm TMF} \rangle$ to the excited states $|n\rangle$\,=$|2\rangle$ to $|16\rangle$ from Table \ref{table:wavefunctions}, associate the observed high-energy spin excitations with the cluster transitions, and identify the corresponding magnetic light scattering mechanism.\cite{fleuryloudon1968} To describe the light-matter interaction in \COSO we use the following effective Raman scattering Hamiltonian:

\begin{eqnarray}
\hat{\mathcal{H}}_{\rm R}=K\sum_{i=1:4}(\hat{S}^{+}_{i}+\hat{S}^{-}_{i}) + \tfrac{1}{2}L\sum_{\substack{i,j=1:4 \\ i\neq j }}\hat{S}^{+}_{i}\hat{S}^{-}_{j} 
\label{eq:ramanhamiltonian}
\end{eqnarray}

The four cluster spin sites are labelled by $i$\,=$1,2,3,4$. The clusters are helimagnetically oriented, with the helimagnetic wave vector $\textbf{q}$ pointing along the [100] equivalent crystallographic axes. \cite{longwavelength} We assume that this ordering leads to a situation where polarization selection rules are always fulfilled for a subset of the spin cluster projections, irrespective of incident and scattered polarization geometry. Thereby, the electric field polarization selection rules are ignored,\cite{fleuryloudon1968} and we only discuss the linear and quadratic spin operators, as indicated in Eq.\,\ref{eq:ramanhamiltonian}. The linear spin operators $\hat{S}^{\pm}_{i}$ ($\Delta S^z$\,$=$\,$\pm 1$) correspond to Elliot-Loudon (one-magnon) scattering. Note that higher order Elliot-Loudon scattering of the form $\hat{S}^{+}_{i}\hat{S}^{+}_{i}$ or $\hat{S}^{-}_{i}\hat{S}^{-}_{i}$ ($\Delta S^z$\,$=$\,$\pm 2$) isn't allowed in \COSO since the spin cluster consists of $S$\,$=$\,$\tfrac{1}{2}$ spins. The exchange scattering (two-magnon) terms are of the form $\hat{S}^{+}_{i}\hat{S}^{-}_{j}$ and $\hat{S}^{-}_{i}\hat{S}^{+}_{j}$ ($\Delta S^z$\,$=$\,$0$),\cite{fleuryloudon1968} but since these operators have the same effect on the cluster wave functions we use the notation $\tfrac{1}{2}\hat{S}^{+}_{i}\hat{S}^{-}_{j}$ in Eq.\,\ref{eq:ramanhamiltonian} to avoid double counting. $K$ and $L$ are arbitrarily valued scattering strengths for the Elliot-Loudon and exchange scattering mechanism, respectively.

The matrix element $M$ for the relevant spin operators $\hat{\mathcal{O}}$ of the form $\hat{S}^{+}_{i}$, $\hat{S}^{-}_{i}$ and $\hat{S}^{+}_{i}\hat{S}^{-}_{j}$ in $\hat{\mathcal{H}}_{\rm R}$ are determined as:

\begin{equation}
M=\vert\langle n\vert \hat{\mathcal{O}} \vert g_{\rm TMF}\rangle\vert^2
\end{equation}

In Table \ref{tb:raman} the spin cluster transitions with nonzero matrix elements and their corresponding Raman modes are indicated. Here, $\vert n\rangle$\,$=$\,$|S,S^z \rangle_{\rm R}$ 
gives the excited cluster state, where it should be understood that $\vert 1\rangle$ is the (isolated cluster) ground state. $\Delta S^z$ indicates the necessary change in spin projection number in order to reach the final state $\vert n\rangle$. When no transition is possible, this is indicated with a hyphen ($-$). All cluster states with a change $\Delta S^z$=$\pm 1, 0$ can be reached either by a one-magnon (Elliot-Loudon) or two-magnon (exchange) scattering process. All the linear and quadratic Raman spin operators which allow for a cluster transition are indicated the table. $E_{\rm A}$ and $E_{\rm B}$ give the transition energies as calculated by spin wave theory.\cite{romhanyi2014,ozerov2014} The final state degeneracy is indicated in between brackets. The second-last column gives the measured Raman shifts $E_{\rm R}$. M$_1$ to M$_6$ refers to the observed spin cluster excitations. The last column gives the cluster excitation energies $E_{\rm ESR}$ observed by electron spin resonance (ESR), as reported in Ref.\,\citenum{ozerov2014}. All energies are indicated in wavenumbers (cm$^{-1}$). 

The $86$\,cm$^{-1}$ M$_{1}$, $204$\,cm$^{-1}$ M$_{2}$, $263$\,cm$^{-1}$ M$_{3}$, and $425$\,cm$^{-1}$ M$_{6}$ modes can be unambiguously identified with different cluster transitions. All these modes are Raman-active through the Elliot-Loudon scattering mechanism ($\hat{S}^{\pm}$ terms). Whether the $273$\,cm$^{-1}$ M$_{4}$ Raman-mode has a $\Delta S^z$\,=\,$0$ or $\pm 1$ cannot be unambiguously identified based on the Raman data set alone. However, this mode has been observed by electron spin resonance, which shows that the mode has a $\Delta S^z$\,=\,$-1$ field behaviour. The Raman-activity of this mode thus also originates from the Elliot-Loudon mechanism. The $300$\,cm$^{-1}$ M$_3$ Raman-mode may either be a transition to $\vert 2,0\rangle_{\rm A_1}$ by a $\hat{S}^{-}$ scattering process, or $\vert 2,1\rangle_{\rm A_1}$ by a $\hat{S}^{+}\hat{S}^{-}$ scattering process. This mode was not observed in the ESR study (indicated with an X). The scattering mechanism of the $300$\,cm$^{-1}$ M$_5$ mode can thus not be unambiguously defined given the present data set and previous works. Scattering to the $\vert 15\rangle$ or $\vert 16\rangle$ excited state (expected excitation energy $\approx$\,$400$\,cm$^{-1}$) is Raman-allowed. This mode however could not be unambiguously resolved due to the presence of a strong phonon mode. For completeness we've indicated the $0$\,cm$^{-1}$ M$_{\rm G}$ in the table, which is the magnetic Goldstone mode of \COSO. The M$_{\rm G}$-mode was observed in the electron spin resonance study. All discussed modes are indicated in Table \ref{tb:raman}.

\begin{table*}
\caption{Magnetic Raman modes, spin cluster excitation energies, and Raman operators. $\vert n\rangle$ and $\vert S,S^z\rangle_{\rm R}$ indicate the final state. The full wavefunctions were given in Table \ref{table:wavefunctions}. The column $\Delta S^z$ gives the change in total cluster spin projection number. Raman operators which allow a specific scattering process between the ground state $\vert g_{\rm TMF}\rangle$, and excited states $\vert 2\rangle$ to $\vert 16\rangle$, are indicated in the columns 1-magnon and 2-magnon. One-magnon scattering is only possible via the Elliot-Loudon mechanism ($\hat{S}^{\pm}$ terms). In the case of \COSO, only the exchange scattering type of 2-magnon scattering is possible, as indicated by Raman operators of a $\hat{S}^{+}\hat{S}^{-}$ form. When no transition between the cluster ground state and excited state is allowed by the respective Raman operator, this is indicated with a hyphen ($-$). $E_{\rm A}$ and $E_{\rm B}$ give the spin wave theory calculated transition energies, as obtained in the work of Refs.\,\citenum{ozerov2014} and \citenum{romhanyi2014}. The degeneracy of the final state is indicated in between brackets. The second-last column gives the measured Raman shifts E$_{\rm R}$. The last column gives the cluster excitation energies $E_{\rm ESR}$ observed by electron spin resonance (ESR), as reported in Ref.\,\citenum{ozerov2014}. All energies are indicated in wavenumbers (cm$^{-1}$).
The Raman-modes at $263$\,cm$^{-1}$ (M$_3$) and $425$\,cm$^{-1}$ (M$_6$) can be unambiguously identified with an $\hat{S}^{-}$ and $\hat{S}^{+}$ transition, which are Raman-active through the Elliot-Loudon scattering mechanism. The weak modes $86$\,cm$^{-1}$ (M$_{1}$) and $204$\,cm$^{-1}$ (M$_{4}$) can be also identified with $\hat{S}^{-}$ and $\hat{S}^{+}$ transitions, respectively. Combined with the ESR result, the $273$\,cm$^{-1}$ M$_4$ Raman-mode can be identified as a transition to $\vert 0,0\rangle_{\rm E}$ via a $\hat{S}^{-}$ operator. The $300$\,cm$^{-1}$ M$_5$ Raman-mode may either be a transition to $\vert 2,0\rangle_{\rm A_1}$ via $\hat{S}^{-}$, or $\vert 2,1\rangle_{\rm A_1}$ via $\hat{S}^{+}\hat{S}^{-}$ terms. This mode was not observed in the ERS study (indicated with an X). The Raman-mode corresponding to a transition to $\vert 1,1\rangle_{\rm E}$ (expected excitation energy $\approx$\,$400$\,cm$^{-1}$) could not be unambiguously identified (indicated with an X). The $0$\,cm$^{-1}$ M$_{\rm G}$ mode corresponds to the magnetic Goldstone mode, and was observed in the ESR-study (Ref.\,\citenum{ozerov2014}).}
\def\arraystretch{1.4}
\small
\begin{tabular}{|c|c|c|c|c|c|c|c|c|}
\hline 
\multirow{2}{*}{$\vert n\rangle$} & final state  & \multirow{2}{*}{$\Delta S^z$} & 1-magnon & 2-magnon & \multirow{2}{*}{${E_1[cm^{-1}]}$} & \multirow{2}{*}{${E_2[cm^{-1}]}$} & \multirow{2}{*}{${E_{\rm R}[cm^{-1}]}$} & \multirow{2}{*}{${E_{\rm ESR}[cm^{-1}]}$}\tabularnewline
 &  $\vert S,S^z\rangle$ &  & (Elliot-Loudon) & (exchange scattering) &  &  &  & \tabularnewline
\hline 
\hline 
$\vert 1\rangle$ & $\vert 1,1\rangle_{\rm A_1}$ & $0$ & $-$ & $-$ & $-$ & $-$ & $-$ & $-$\tabularnewline
\hline 
\multirow{2}{*}{$\vert 2\rangle$} & \multirow{2}{*}{$\vert 1,0\rangle_{\rm A_1}$} & \multirow{2}{*}{$-1$} & \multirow{2}{*}{$\hat{S}^{-}_{1}$,$\hat{S}^{-}_{2}$,$\hat{S}^{-}_{3}$,$\hat{S}^{-}_{4}$} & \multirow{2}{*}{$-$} & \multirow{2}{*}{$0(\textit{1})$} & \multirow{2}{*}{$96(\textit{3})$} & \multirow{2}{*}{$85({\rm M}_{1})$} & $0({\rm M}_{\rm G})$\tabularnewline
 &  &  &  &  &  &  &  & $85({\rm M}_{1})$\tabularnewline
\hline 
$\vert 3\rangle$ & $\vert 1,\bar{1}\rangle_{\rm A_1}$ & $-2$ & $-$ & $-$ & $159(\textit{4})$ & $-$ & $-$ & $-$\tabularnewline
\hline 
$\vert 4\rangle$ & $\vert 0,0\rangle_{\rm E_1}$ & $-1$ & $\hat{S}^{-}_{_2}$,$\hat{S}^{-}_{3}$,$\hat{S}^{-}_{4}$ & $-$ & \multirow{2}{*}{$236(\textit{3})$} & \multirow{2}{*}{$276(\textit{5})$} & \multirow{2}{*}{$273({\rm M}_{4})$} & \multirow{2}{*}{$270({\rm M}_{4})$}\tabularnewline
\cline{1-5} 
$\vert 5\rangle$ & $\vert 0,0\rangle_{\rm E_2}$ & $-1$ & $\hat{S}^{-}_{3}$,$\hat{S}^{-}_{4}$ & $-$ &  &  &  & \tabularnewline
\hline 
$\vert 6\rangle$ & $\vert 2,\bar{2}\rangle_{\rm A_1}$ & $-3$ & $-$ & $-$ & $338(\textit{4})$ & $-$ & $-$ & $-$\tabularnewline
\hline 
$\vert 7\rangle$ & $\vert 2,\bar{1}\rangle_{\rm A_1}$ & $-2$ & $-$ & $-$ & $335(\textit{4})$ & $-$ & $-$ & $-$\tabularnewline
\hline 
$\vert 8\rangle$ & $\vert 2,0\rangle_{\rm A_1}$ & $-1$ & $\hat{S}^{-}_{1}$,$\hat{S}^{-}_{2}$,$\hat{S}^{-}_{3}$,$\hat{S}^{-}_{4}$ & $-$ & $310(\textit{1})$ & $314(\textit{3})$ & $300({\rm M}_{5}?)$ & X\tabularnewline
\hline 
\multirow{3}{*}{$\vert 9\rangle$} & \multirow{3}{*}{$\vert 2,1\rangle_{\rm A_1}$} & \multirow{3}{*}{$0$} & \multirow{3}{*}{$-$} & $\hat{S}^{+}_{1}\hat{S}^{-}_{2}$,$\hat{S}^{+}_{1}\hat{S}^{-}_{3}$,$\hat{S}^{+}_{1}\hat{S}^{-}_{4}$,$\hat{S}^{+}_{2}\hat{S}^{-}_{1}$, & \multirow{3}{*}{$272(\textit{1})$} & \multirow{3}{*}{$288(\textit{3})$} & \multirow{3}{*}{$300({\rm M}_{5}?)$} & \multirow{3}{*}{$-$}\tabularnewline
 &  &  &  & $\hat{S}^{+}_{2}\hat{S}^{-}_{3}$,$\hat{S}^{+}_{2}\hat{S}^{-}_{4}$,$\hat{S}^{+}_{3}\hat{S}^{-}_{1}$,$\hat{S}^{+}_{3}\hat{S}^{-}_{2}$, &  &  &  & \tabularnewline
 &  &  &  & $\hat{S}^{+}_{3}\hat{S}^{-}_{4}$,$\hat{S}^{+}_{4}\hat{S}^{-}_{1}$,$\hat{S}^{+}_{4}\hat{S}^{-}_{2}$,$\hat{S}^{+}_{4}\hat{S}^{-}_{3}$ &  &  &  & \tabularnewline
\hline 
\multirow{2}{*}{$\vert 10\rangle$} & \multirow{2}{*}{$\vert 2,2\rangle_{\rm A_1}$} & \multirow{2}{*}{$+1$} & \multirow{2}{*}{$\hat{S}^{+}_{1}$,$\hat{S}^{+}_{2}$,$\hat{S}^{+}_{3}$,$\hat{S}^{+}_{4}$} & \multirow{2}{*}{$-$} & \multirow{2}{*}{$206(\textit{1})$} & \multirow{2}{*}{$260(\textit{3})$} & $204({\rm M}_{2})$ & $202({\rm M}_{2})$\tabularnewline
 &  &  &  &  &  &  & $263({\rm M}_{3})$ & $263({\rm M}_{3})$\tabularnewline
\hline 
$\vert 11\rangle$ & $\vert 1,\bar{1}\rangle_{\rm E_1}$ & $-2$ & $-$ & $-$ & \multirow{2}{*}{$380(\textit{8})$} & \multirow{2}{*}{$-$} & \multirow{2}{*}{$-$} & \multirow{2}{*}{$-$}\tabularnewline
\cline{1-5} 
$\vert 12\rangle$ & $\vert 1,\bar{1}\rangle_{\rm E_2}$ & $-2$ & $-$ & $-$ &  &  &  & \tabularnewline
\hline 
$\vert 13\rangle$ & $\vert 1,0\rangle_{\rm E_1}$ & $-1$ & $\hat{S}^{-}_{2}$,$\hat{S}^{-}_{3}$,$\hat{S}^{-}_{4}$ & $-$ & \multirow{2}{*}{$419(\textit{8})$} & \multirow{2}{*}{$-$} & \multirow{2}{*}{$425({\rm M}_{6})$} & \multirow{2}{*}{$420({\rm M}_{6})$}\tabularnewline
\cline{1-5} 
$\vert 14\rangle$ & $\vert 1,0\rangle_{\rm E_2}$ & $-1$ & $\hat{S}^{-}_{3}$,$\hat{S}^{-}_{4}$ & $-$ &  &  &  & \tabularnewline
\hline 
\multirow{3}{*}{$\vert 15\rangle$} & \multirow{3}{*}{$\vert 1,1\rangle_{\rm E_1}$} & \multirow{3}{*}{$0$} & \multirow{3}{*}{$-$} & $\hat{S}^{+}_{1}\hat{S}^{-}_{2}$,$\hat{S}^{+}_{1}\hat{S}^{-}_{3}$,$\hat{S}^{+}_{1}\hat{S}^{-}_{4}$, & \multirow{5}{*}{$396(\textit{3})$} & \multirow{5}{*}{$401(\textit{5})$} & \multirow{5}{*}{X} & \multirow{5}{*}{$-$}\tabularnewline
 &  &  &  & $\hat{S}^{+}_{2}\hat{S}^{-}_{3}$,$\hat{S}^{+}_{2}\hat{S}^{-}_{4}$,$\hat{S}^{+}_{3}\hat{S}^{-}_{1}$, &  &  &  & \tabularnewline
 &  &  &  & $\hat{S}^{+}_{3}\hat{S}^{-}_{4}$,$\hat{S}^{+}_{4}\hat{S}^{-}_{2}$,$\hat{S}^{+}_{4}\hat{S}^{-}_{3}$ &  &  &  & \tabularnewline
\cline{1-5} 
\multirow{2}{*}{$\vert 16\rangle$} & \multirow{2}{*}{$\vert 1,1\rangle_{\rm E_2}$} & \multirow{2}{*}{$0$} & \multirow{2}{*}{$-$} & $\hat{S}^{+}_{1}\hat{S}^{-}_{3}$,$\hat{S}^{+}_{1}\hat{S}^{-}_{4}$,$\hat{S}^{+}_{2}\hat{S}^{-}_{3}$, &  &  &  & \tabularnewline
 &  &  &  & $\hat{S}^{+}_{2}\hat{S}^{-}_{4}$,$\hat{S}^{+}_{3}\hat{S}^{-}_{4}$,$\hat{S}^{+}_{4}\hat{S}^{-}_{3}$ &  &  &  & \tabularnewline
\hline 
\end{tabular}
\label{tb:raman}
\end{table*}

\subsection{Temperature dependence of spin cluster excitations}

\begin{figure}[h]
 \center
\includegraphics[width=3.375in]{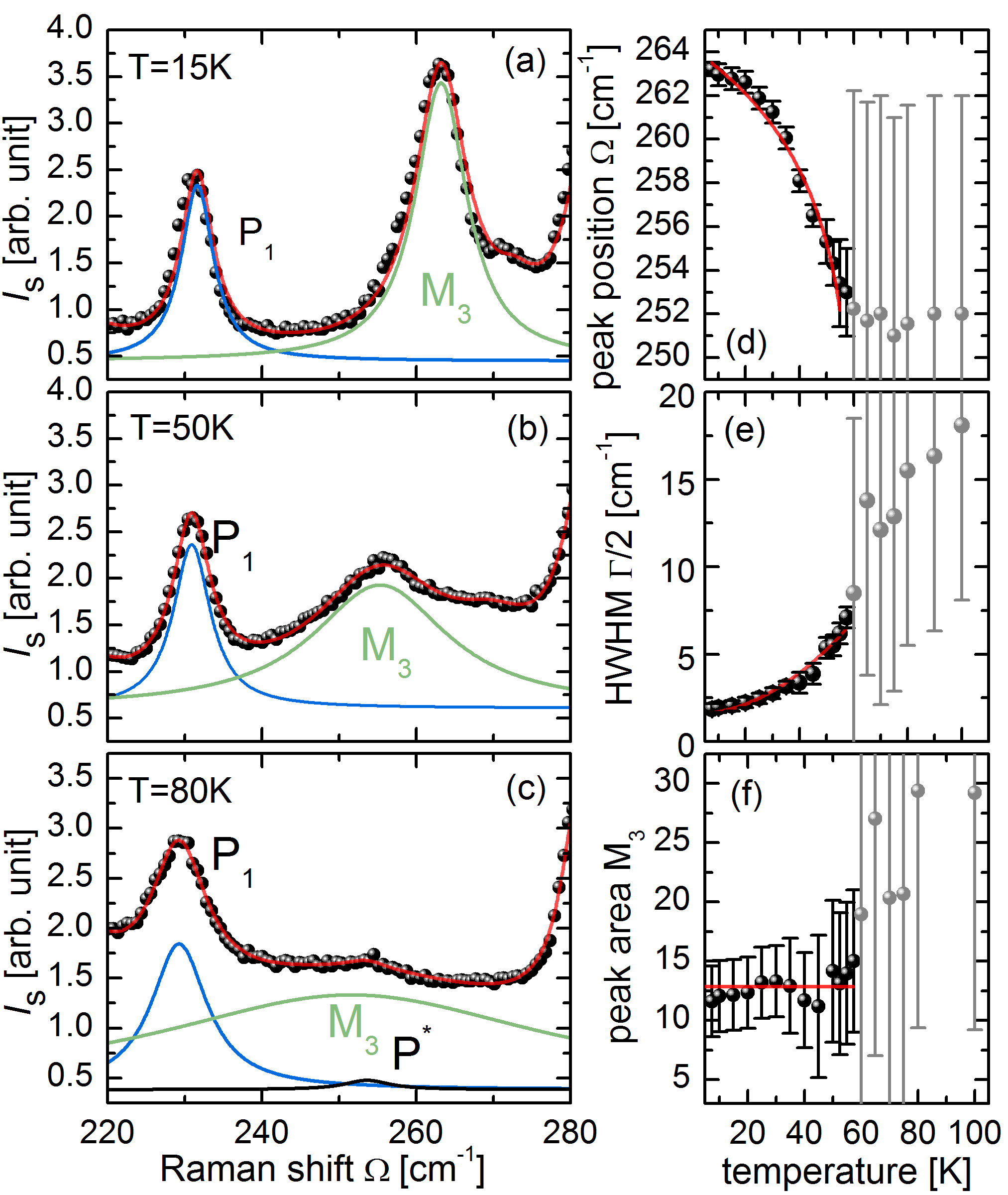} 
\caption{(a-c) Scattering region around the M$_3$ spin cluster mode and the P$_1$ phonon for temperatures below and above $T_{\rm C}$. (d) Temperature dependence of the M$_3$ spin excitation energy. Below $T_{\rm C}$ the temperature dependent position is well-fitted with a scaling function of the form $\Omega$(T)$\propto$\,$\big(T_C-T)/T_C \big)^{\gamma}$, with $\gamma$\,$\approx$\,$0.02$ as exponent (red curve). (e) Temperature dependent spectral weight of the M$_3$ spin excitation. Below $T_{\rm C}$\,$\approx$\,$58$\,K the spectral weight remains constant (red curve). Above $T_{\rm C}$\,$\approx$\,$58$\,K the determination of the spectral weight becomes unreliable, however, it stays finite. (f) Half width at half maximum (inverse decay rate) of the M$_1$ spin excitation. Below $T_{\rm C}$ a quadratic power law is observed, in addition to a strong spontaneous decay rate.}
\label{fig:mode265}
\end{figure}

For the spin cluster excitations a spectral weight transfer to lower Raman shift $\Omega$ is observed when the temperature increases towards $T_{\rm C}$ (see Fig.\,\ref{fig:magnonraman}). The magnetic spectral weight transfer is understood as a softening and broadening of the high-energy spin excitations. In Fig.\,\ref{fig:mode265}a-c we shown the scattering region around the M$_3$ spin cluster mode and the P$_1$ phonon in closer detail for temperatures below and above $T_{\rm C}$. The scattering region is fitted with a sum of Lorentzian fit functions. The red line shows the full fit and the blue and green lines the fits for the P$_1$ phonon and M$_3$ spin excitation, respectively. Above $T_{\rm C}$ the M$_3$ excitation has significantly broadened and a weak phonon becomes visible (P$^*$). The broadening above $T_{\rm C}$ has previously been identified in the terahertz transmission of \COSO.\cite{laurita2017} 

The frequency $\Omega$($T$) of the M$_3$ excitation is plotted in Fig.\,\ref{fig:mode265}d. Below $T_{\rm C}$ the temperature dependence of the excitation energy can be well-described by a scaling law of the form $\Omega$($T$)$\propto$\,$\big(T_{\rm C}-T)/T_{\rm C} \big)^{\gamma}$, with $\gamma$\,$\approx$\,$0.02$ as exponent. Above $T_{\rm C}$\,$\approx$\,$58$\,K magnetic scattering still persists, but the spin excitations have significantly broadened into a continuum type of magnetic scattering.\cite{kurnosov2012} This is most clearly seen in Fig.\,\ref{fig:mode265}c for M$_3$ ($\Delta S^z$\,=\,$+$\,$1$). For M$_4$ ($\Delta S^z$\,=\,$-$\,$1$) similar qualitative behaviour is observed, as most clearly seen in Fig.\,\ref{fig:magnonraman}. Above $T_{\rm C}$ the fitting of the M$_3$ peak position becomes unreliable. The temperature dependent spectral weight of M$_3$ is plotted in Fig.\,\ref{fig:mode265}f. Up till $T_{\rm C}$ the spectral weight remains constant. Above $T_{\rm C}$ the determination of the spectral weight becomes unreliable.

Figure \ref{fig:mode265}e shows the temperature dependence of the line width at half maximum (inverse decay rate) of the M$_3$ excitation. The functional temperature dependence below $T_{\rm C}$ is well fitted by the second order polynomial $\frac{\Gamma}{2}(T)=\frac{\Gamma_0}{2}(T=0) + A\cdot T + B\cdot T^2$, with the largest contributions formed by $\frac{\Gamma_0}{2}(T=0)$ and $B\cdot T^2$. The latter process describes a four magnon interaction.\cite{laurita2017} The large finite $\frac{\Gamma_0}{2}(T=0)$ term may result from inhomogeneous broadening from disorder. However, Laurita \textit{et al.} (Ref.\,\citenum{laurita2017}) argue that the spontaneous decay rate of spin cluster excitations in \COSO instead may originate from quantum fluctuations.

The temperature dependence of the magnetic scattering across the phase transition of the high-energy spin excitations in \COSO  is rather peculiar in light of the vast range of historic and contemporary magnetic Raman scattering literature. \cite{cottam1986light,devereaux2007} In the case of single spin (anti-)ferromagnets, such as NiF$_2$ or KNiF$_3$, the first order (Elliot-Loudon) scattering originates from low-energy zone-\textit{center} magnons. The $\Gamma$-point one-magnon scattering vanishes above the (anti-)ferromagnetic critical temperature $T_{\rm N/C}$, where the long-range spin correlation is strongly reduced. Exchange scattering from high-energy zone-\textit{edge} magnon pairs, described by the pair-operators $\hat{S}^{+}S^{-}$ and $\hat{S}^{-}\hat{S}^{+}$ ($\Delta S^z$\,=\,$0$), is possible in antiferromagnets above $T_{\rm N}$, since short-range correlations still exist. \cite{cottam1986light,fleuryloudon1968} 

\begin{figure}[ht]
 \center
\includegraphics[width=3.375in]{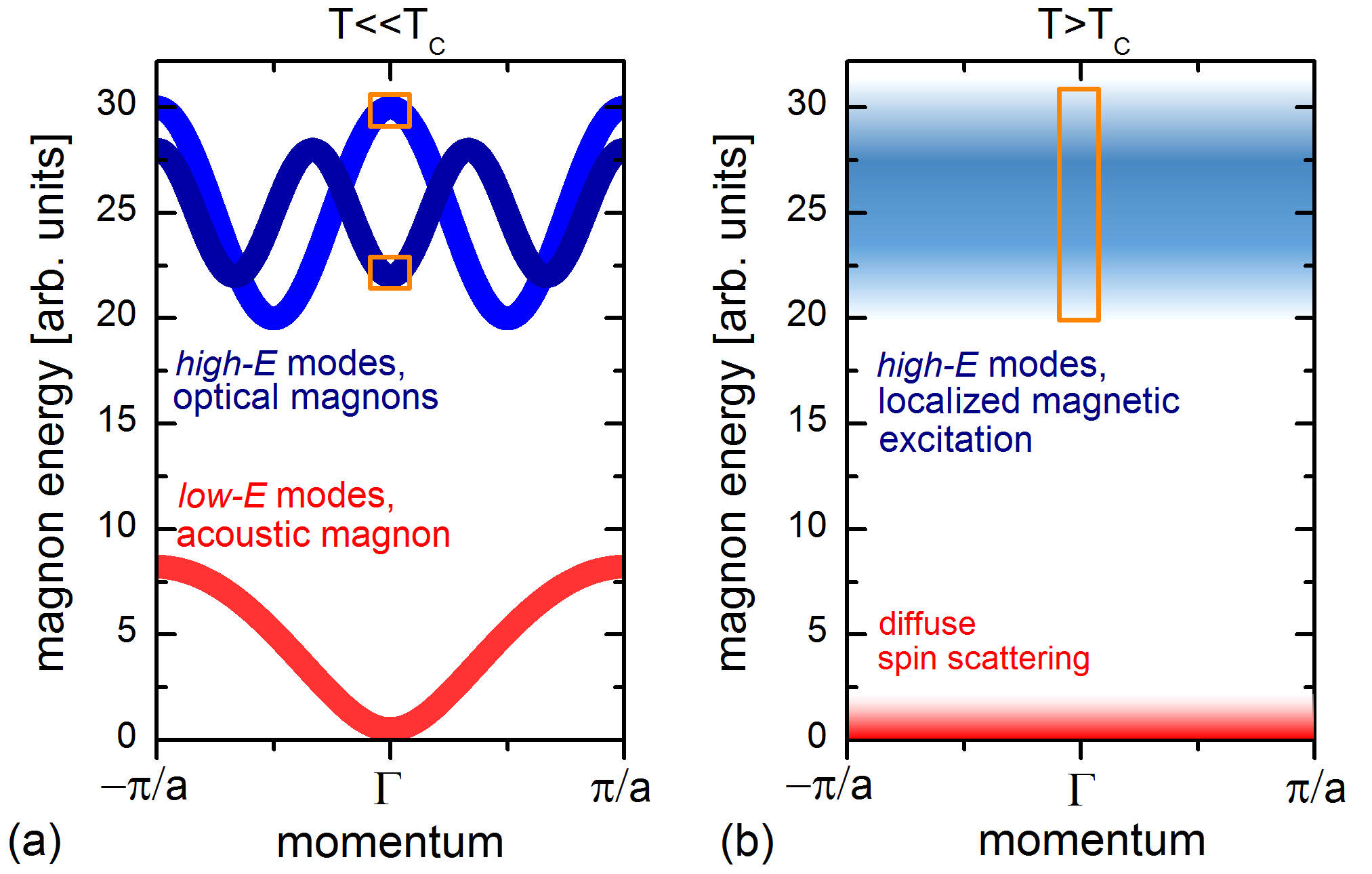} 
\caption{Cartoon picture of the spin excitation dispersion of a cluster magnet below and above the critical temperature $T_{\rm C}$. (a) A cluster magnet has well-defined low-energy \textit{external} spin cluster excitation branches (red) and high energy \textit{internal} spin cluster excitation branches (blue) below $T_{\rm C}$. Raman-scattering from high-energy spin cluster excitations is possible at the $\Gamma$-point, as indicated with orange squares. (b) Above $T_{\rm C}$ the high-energy internal spin cluster excitation branches cross over into a broad dispersionless band of localized magnetic excitations (blue). The low energy external branch vanishes above $T_{\rm C}$ due to the loss of inter-cluster correlations, leading to diffuse spin scattering. Raman-scattering from localized high-energy spin cluster excitations is possible at the $\Gamma$-point, as indicated with the orange rectangle.}
\label{fig:clustermagnons}
\end{figure}

The formation of Cu$_4$ spin clusters far above the long-range ordering temperature $T_{\rm C}$\,$\approx$\,$58$\,K, \cite{tucker2016} and the resulting high-energy dispersive magnon branch below $T_{\rm C}$ results in the possibility to scatter from zone-center \textit{internal spin cluster} excitations above and below $T_{\rm C}$ by the Elliot-Loudon mechanism. \cite{cottam1986light,fleuryloudon1968} Below $T_{\rm C}$ high-energy \textit{optical magnon} branches are well-defined and dispersive by the inter-cluster correlation. A cartoon is provided in Fig.\,\ref{fig:clustermagnons}a. Raman-scattering by the Elliot-Loudon mechanism is possible at the $\Gamma$-point (indicated with orange squares). However, above $T_{\rm C}$ inter-cluster correlations are lost. Here the high-energy spin cluster excitations are thus of fully cluster-internal nature.\cite{liu1976magnetic}  The broad magnetic scattering evidences that the uncoupled clusters reside in an inhomogeneous environment and/or that the lifetime of the cluster-internal spin cluster excitations is short. In reciprocal space this corresponds to a 
broad dispersionless band of localized cluster-internal spin excitations,\cite{liu1976magnetic} as depicted in Fig.\,\ref{fig:clustermagnons}b This finite localized spin cluster excitation density-of states at the $\Gamma$-point above $T_{\rm C}$ still allows for first-order scattering, but however will appear as broad continuum of magnetic scattering, as indicated with the orange rectangle.  \cite{liu1976magnetic} We stress out that it is thus the spin cluster nature with resulting $\Gamma$-point optical magnons, which make the one-magnon excitations of high enough energy to be observable in \COSO. This is in sharp contrast to simple antiferromagnets, where a relatively strong magnetocrystalline anisotropy is necessary to observe one-magnon excitations by Raman spectroscopy.

\subsection{Magnetoelastic coupling}
\COSO is a rare example of a magnetoelectric material with $pd$-hybridization as coupling mechanism.\cite{seki2012} The magnetoelectric coupling for instance allows to control the angular orientation of the skyrmion lattice. \cite{white2014electric,white2018electric}  Different reports addressed that no significant magnetostrictive lattice contraction nor a structural symmetry change occurs in the magnetically ordered phase, \cite{jwbos2008,kurnosov2012} even though the natural optical activity shows an enhancement in the helimagnetic phase.\cite{versteeg2016} This is in line with that \COSO has $pd$-hybridization as the dominant magnetoelectric coupling mechanism. However, this observation does not imply that magnetoelastic coupling is completely absent in \COSO. Evidences of a finite magnetoelastic coupling are, for instance, the anomalies in optical phonon frequencies around $T_{\rm C}$ \cite{gnezdilov2010,miller2010,kurnosov2012}  and the observation that the propagation of acoustic phonons is nonreciprocal in \COSO. \cite{nomura2018phonon}   

\begin{figure}[h!]
 \center
\includegraphics[width=3.375in]{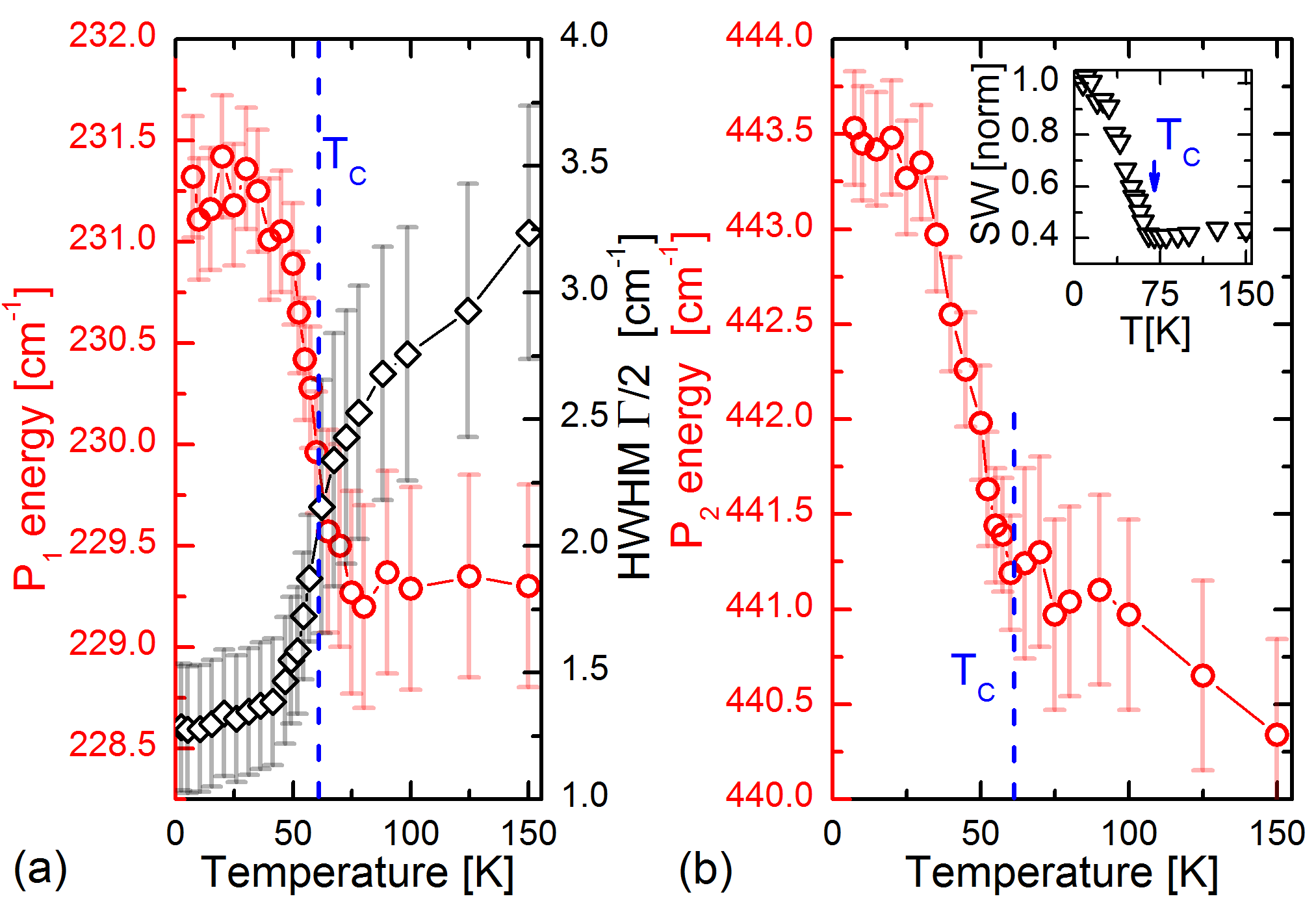}   
\caption{(a) P$_1$ phonon energy and half width at full maximum (HWHM) and (b) P$_2$ phonon energy with the normalized spectral weight SW as inset. Around the magnetic critical temperature $T_{\rm C}$\,$\approx$\,$58$\,K an anomaly is observed in the P$_1$ and P$_2$ phonon parameters.}
\label{fig:phononwidthposition}
\end{figure}

In Fig. \ref{fig:phononwidthposition}a we plot the phonon energy and half width at half maximum (HWHM) for the phonon P$_1$. Figure \ref{fig:phononwidthposition}a shows the phonon energy and spectral weight (SW) for the phonon P$_2$. All plotted phonon parameters show strong sensitivity to magnetic ordering. This is especially the case for the P$_2$ phonon, as for instance seen from the spectral weight, but also directly in Fig.\,\ref{fig:magnonraman}. Neither, the spin wave theory calculations, nor neutron experiments evidence the presence of a spin cluster excitation around $444$\,cm$^{-1}$. Instead, the P$_2$ mode corresponds to the vibration of the CuO$_5$ pyramidal units. \cite{miller2010} The similar energy scale and overlapping dispersion of optical phonons and high-energy spin cluster excitations can lead to a phonon-magnon hybridization by magnetoelastic coupling.\cite{kittel1958,shen1966,nomura2018phonon} This in turn will lead to a strong temperature dependence for the line-width, position and spectral weight of the optical phonons.

\section{Conclusions}
A Raman spectroscopy study of the cluster Mott insulator \COSO was performed. Multiple high-energy spin cluster excitations were observed besides a rich phonon spectrum. We systematically characterized the observed spin cluster transitions along the lines of the \COSO spin cluster model and deduced that the Raman activity of the spin cluster excitations originates in the first order Elliot-Loudon light scattering mechanism. The high energy spin cluster excitation modes show to soften and broaden with increasing temperature and persist above $T_{\rm C}$ as a broad magnetic scattering continuum. Above $T_{\rm C}$, the Cu$_4$ clusters are decoupled, resulting in localized cluster-internal spin excitations. In the long-range ordered phase the internal cluster modes acquire dispersion by the inter-cluster exchange  interactions, and form optical magnon branches, resulting in well-defined magnetic modes in the Raman spectrum. Our observations support the picture that \COSO can be regarded as a solid-state molecular crystal of spin nature. 

\section*{Acknowledgments}
This project was partially financed by the Deutsche Forschungsgemeinschaft (DFG) through Project No. 277146847 - Collaborative Research Center 1238:
Control and Dynamics of Quantum Materials (Subprojects No. A02 and No. B03). RBV acknowledges funding through the Bonn-Cologne Graduate School of Physics and Astronomy (BCGS). All authors thank D.~I.~Khomskii (Cologne, DE) and L. Bohat\'{y} (Cologne, DE) for fruitful discussions. RBV thanks D.~Inosov (Dresden, DE), I.~Rousochatzakis (Loughborough University, UK) and J.~Romh\'{a}nyi
(Okinawa, JP) for fruitful discussions on their \COSO spin cluster publications.


\end{document}